\documentclass[aps,prd,twocolumn,superscriptaddress,showpacs,amsmath,amssymb]{revtex4}

\pdfoutput=1

\usepackage{graphicx}

\usepackage{dcolumn}
\usepackage{bm}

\newcommand{\missEt}{\ensuremath{ \not\!\!E_T  } }
\newcommand{\JES}{\Delta_{\mathrm{JES}}}
\newcommand{\Cs}{\nu_{\mathrm{sig}}}

\newcommand{\Wa}{W_A}
\newcommand{\Wj}{W_E}
\newcommand{\Et}{E_T}

\newcommand{\IntLum}{3.2~fb$^{-1}$}
\newcommand{\NcResult}{578}

\begin{document}


\title{Measurement of the top-quark mass in the lepton+jets channel using a matrix element technique  with the   CDF II detector}


\affiliation{}


\affiliation{Institute of Physics, Academia Sinica, Taipei, Taiwan 11529, Republic of China} 
\affiliation{Argonne National Laboratory, Argonne, Illinois 60439, USA} 
\affiliation{University of Athens, 157 71 Athens, Greece} 
\affiliation{Institut de Fisica d'Altes Energies, ICREA, Universitat Autonoma de Barcelona, E-08193, Bellaterra (Barcelona), Spain} 
\affiliation{Baylor University, Waco, Texas 76798, USA} 
\affiliation{Istituto Nazionale di Fisica Nucleare Bologna, $^{aa}$University of Bologna, I-40127 Bologna, Italy} 
\affiliation{University of California, Davis, Davis, California 95616, USA} 
\affiliation{University of California, Los Angeles, Los Angeles, California 90024, USA} 
\affiliation{Instituto de Fisica de Cantabria, CSIC-University of Cantabria, 39005 Santander, Spain} 
\affiliation{Carnegie Mellon University, Pittsburgh, Pennsylvania 15213, USA} 
\affiliation{Enrico Fermi Institute, University of Chicago, Chicago, Illinois 60637, USA}
\affiliation{Comenius University, 842 48 Bratislava, Slovakia; Institute of Experimental Physics, 040 01 Kosice, Slovakia} 
\affiliation{Joint Institute for Nuclear Research, RU-141980 Dubna, Russia} 
\affiliation{Duke University, Durham, North Carolina 27708, USA} 
\affiliation{Fermi National Accelerator Laboratory, Batavia, Illinois 60510, USA} 
\affiliation{University of Florida, Gainesville, Florida 32611, USA} 
\affiliation{Laboratori Nazionali di Frascati, Istituto Nazionale di Fisica Nucleare, I-00044 Frascati, Italy} 
\affiliation{University of Geneva, CH-1211 Geneva 4, Switzerland} 
\affiliation{Glasgow University, Glasgow G12 8QQ, United Kingdom} 
\affiliation{Harvard University, Cambridge, Massachusetts 02138, USA} 
\affiliation{Division of High Energy Physics, Department of Physics, University of Helsinki and Helsinki Institute of Physics, FIN-00014, Helsinki, Finland} 
\affiliation{University of Illinois, Urbana, Illinois 61801, USA} 
\affiliation{The Johns Hopkins University, Baltimore, Maryland 21218, USA} 
\affiliation{Institut f\"{u}r Experimentelle Kernphysik, Karlsruhe Institute of Technology, D-76131 Karlsruhe, Germany} 
\affiliation{Center for High Energy Physics: Kyungpook National University, Daegu 702-701, Korea; Seoul National University, Seoul 151-742, Korea; Sungkyunkwan University, Suwon 440-746, Korea; Korea Institute of Science and Technology Information, Daejeon 305-806, Korea; Chonnam National University, Gwangju 500-757, Korea; Chonbuk National University, Jeonju 561-756, Korea} 
\affiliation{Ernest Orlando Lawrence Berkeley National Laboratory, Berkeley, California 94720, USA} 
\affiliation{University of Liverpool, Liverpool L69 7ZE, United Kingdom} 
\affiliation{University College London, London WC1E 6BT, United Kingdom} 
\affiliation{Centro de Investigaciones Energeticas Medioambientales y Tecnologicas, E-28040 Madrid, Spain} 
\affiliation{Massachusetts Institute of Technology, Cambridge, Massachusetts 02139, USA} 
\affiliation{Institute of Particle Physics: McGill University, Montr\'{e}al, Qu\'{e}bec, Canada H3A~2T8; Simon Fraser University, Burnaby, British Columbia, Canada V5A~1S6; University of Toronto, Toronto, Ontario, Canada M5S~1A7; and TRIUMF, Vancouver, British Columbia, Canada V6T~2A3} 
\affiliation{University of Michigan, Ann Arbor, Michigan 48109, USA} 
\affiliation{Michigan State University, East Lansing, Michigan 48824, USA}
\affiliation{Institution for Theoretical and Experimental Physics, ITEP, Moscow 117259, Russia}
\affiliation{University of New Mexico, Albuquerque, New Mexico 87131, USA} 
\affiliation{Northwestern University, Evanston, Illinois 60208, USA} 
\affiliation{The Ohio State University, Columbus, Ohio 43210, USA} 
\affiliation{Okayama University, Okayama 700-8530, Japan} 
\affiliation{Osaka City University, Osaka 588, Japan} 
\affiliation{University of Oxford, Oxford OX1 3RH, United Kingdom} 
\affiliation{Istituto Nazionale di Fisica Nucleare, Sezione di Padova-Trento, $^{bb}$University of Padova, I-35131 Padova, Italy} 
\affiliation{LPNHE, Universite Pierre et Marie Curie/IN2P3-CNRS, UMR7585, Paris, F-75252 France} 
\affiliation{University of Pennsylvania, Philadelphia, Pennsylvania 19104, USA}
\affiliation{Istituto Nazionale di Fisica Nucleare Pisa, $^{cc}$University of Pisa, $^{dd}$University of Siena and $^{ee}$Scuola Normale Superiore, I-56127 Pisa, Italy} 
\affiliation{University of Pittsburgh, Pittsburgh, Pennsylvania 15260, USA} 
\affiliation{Purdue University, West Lafayette, Indiana 47907, USA} 
\affiliation{University of Rochester, Rochester, New York 14627, USA} 
\affiliation{The Rockefeller University, New York, New York 10065, USA} 
\affiliation{Istituto Nazionale di Fisica Nucleare, Sezione di Roma 1, $^{ff}$Sapienza Universit\`{a} di Roma, I-00185 Roma, Italy} 

\affiliation{Rutgers University, Piscataway, New Jersey 08855, USA} 
\affiliation{Texas A\&M University, College Station, Texas 77843, USA} 
\affiliation{Istituto Nazionale di Fisica Nucleare Trieste/Udine, I-34100 Trieste, $^{gg}$University of Udine, I-33100 Udine, Italy} 
\affiliation{University of Tsukuba, Tsukuba, Ibaraki 305, Japan} 
\affiliation{Tufts University, Medford, Massachusetts 02155, USA} 
\affiliation{University of Virginia, Charlottesville, Virginia 22906, USA}
\affiliation{Waseda University, Tokyo 169, Japan} 
\affiliation{Wayne State University, Detroit, Michigan 48201, USA} 
\affiliation{University of Wisconsin, Madison, Wisconsin 53706, USA} 
\affiliation{Yale University, New Haven, Connecticut 06520, USA} 
\author{T.~Aaltonen}
\affiliation{Division of High Energy Physics, Department of Physics, University of Helsinki and Helsinki Institute of Physics, FIN-00014, Helsinki, Finland}
\author{B.~\'{A}lvarez~Gonz\'{a}lez$^w$}
\affiliation{Instituto de Fisica de Cantabria, CSIC-University of Cantabria, 39005 Santander, Spain}
\author{S.~Amerio}
\affiliation{Istituto Nazionale di Fisica Nucleare, Sezione di Padova-Trento, $^{bb}$University of Padova, I-35131 Padova, Italy} 

\author{D.~Amidei}
\affiliation{University of Michigan, Ann Arbor, Michigan 48109, USA}
\author{A.~Anastassov}
\affiliation{Northwestern University, Evanston, Illinois 60208, USA}
\author{A.~Annovi}
\affiliation{Laboratori Nazionali di Frascati, Istituto Nazionale di Fisica Nucleare, I-00044 Frascati, Italy}
\author{J.~Antos}
\affiliation{Comenius University, 842 48 Bratislava, Slovakia; Institute of Experimental Physics, 040 01 Kosice, Slovakia}
\author{G.~Apollinari}
\affiliation{Fermi National Accelerator Laboratory, Batavia, Illinois 60510, USA}
\author{J.A.~Appel}
\affiliation{Fermi National Accelerator Laboratory, Batavia, Illinois 60510, USA}
\author{A.~Apresyan}
\affiliation{Purdue University, West Lafayette, Indiana 47907, USA}
% ...
\collaboration{CDF Collaboration\footnote{With visitors from $^a$Istituto Nazionale di Fisica Nucleare, Sezione di Cagliari, 09042 Monserrato (Cagliari), Italy,
$^b$University of CA Irvine, Irvine, CA  92697, USA,
$^c$University of CA Santa Barbara, Santa Barbara, CA 93106, USA,
$^d$University of CA Santa Cruz, Santa Cruz, CA  95064, USA,
$^e$CERN,CH-1211 Geneva, Switzerland,
$^f$Cornell University, Ithaca, NY  14853, USA, 
$^g$University of Cyprus, Nicosia CY-1678, Cyprus, 
$^h$Office of Science, U.S. Department of Energy, Washington, DC 20585, USA,
$^i$University College Dublin, Dublin 4, Ireland,
$^j$University of Fukui, Fukui City, Fukui Prefecture, Japan 910-0017,
$^k$Universidad Iberoamericana, Mexico D.F., Mexico,
$^l$Iowa State University, Ames, IA  50011, USA,
$^m$University of Iowa, Iowa City, IA  52242, USA,
$^n$Kinki University, Higashi-Osaka City, Japan 577-8502,
$^o$Kansas State University, Manhattan, KS 66506, USA,
$^p$University of Manchester, Manchester M13 9PL, United Kingdom,
$^q$Queen Mary, University of London, London, E1 4NS, United Kingdom,
$^r$University of Melbourne, Victoria 3010, Australia,
$^s$Muons, Inc., Batavia, IL 60510, USA,
$^t$Nagasaki Institute of Applied Science, Nagasaki, Japan, 
$^u$National Research Nuclear University, Moscow, Russia,
$^v$University of Notre Dame, Notre Dame, IN 46556, USA,
$^w$Universidad de Oviedo, E-33007 Oviedo, Spain, 
$^x$Texas Tech University, Lubbock, TX  79609, USA,
$^y$Universidad Tecnica Federico Santa Maria, 110v Valparaiso, Chile,
$^z$Yarmouk University, Irbid 211-63, Jordan,
$^{hh}$On leave from J.~Stefan Institute, Ljubljana, Slovenia, 
}}
\noaffiliation

\date{\today}

\begin{abstract}
A measurement of the top-quark mass is presented using Tevatron data
from proton-antiproton collisions at center-of-mass energy
$\sqrt{s}=1.96$ TeV collected with the CDF II detector.  Events are
selected from a sample of candidates for production of $t\bar t$
pairs that decay into the lepton+jets channel. The top-quark mass is
measured with an unbinned maximum likelihood method where the event
probability density functions are calculated using signal and
background matrix elements, as well as a set of parameterized
jet-to-parton transfer functions.  The likelihood function is
maximized with respect to the top-quark mass, the signal fraction in
the sample, and a correction to the jet energy scale (JES) calibration
of the calorimeter jets. The simultaneous measurement of the JES
correction ($\JES$) amounts to an additional \textit{in situ} jet
energy calibration based on the known mass of the hadronically
decaying $W$ boson.  Using the data sample of \NcResult\ lepton+jets
candidate events, corresponding to \IntLum\ of integrated luminosity,
the top-quark mass is measured to be $m_t = \rm 172.4 \pm
1.4\:(stat+\JES) \pm 1.3\:(syst)~GeV/{\it c}^2$.
\end{abstract}


\maketitle

\clearpage

The top-quark mass, $m_t$, is an intrinsic parameter of the standard
model (SM) of particle physics and is of particular importance due to
its strikingly large value.  As a result, the top quark has a large
effect on radiative corrections to electroweak processes and has a
Yukawa coupling to the Higgs field of $\mathcal{O}(1)$, which may
provide insight into the mechanism of electroweak symmetry
breaking~\cite{EWSBreview}.

The Higgs boson mass, $m_H$, is not predicted by the SM, but
constraints on its value can be derived from the calculation of
radiative corrections to the $W$~boson mass, $m_W$, and from the
values of other precision electroweak variables~\cite{LEPcomb}.
These corrections depend primarily on $\ln m_H$ and $m_t^2$, and thus
precision measurements of $m_W$ and $m_t$ provide important
constraints on $m_H$.

The dominant top-quark production process is pair production via the
strong interaction. At Fermilab's Tevatron this process is initiated
by $p\bar{p}$ collisions at center-of-mass energy $\sqrt{s}=\rm
1.96~TeV$.  Because of its large mass, the top quark decays rapidly
with lifetime $\tau_t \sim 10^{-25}$~s~\cite{top_width} --- fast
enough that it has essentially no time to interact and may be
considered as a free quark. This allows a direct measurement of its
mass from the daughter particles from its decay, and as a result $m_t$
has the lowest relative uncertainty of all of the quark
masses~\cite{pdg}.

In the SM top quarks decay via the weak interaction, predominantly to
$W$~bosons and $b$~quarks as \linebreak $t\bar{t} \to W^+ b \: W^- \bar{b}$.
$W$~bosons decay into lower-mass fermion-antifermion pairs: a charged
lepton and a neutrino ($W^+ \to \bar{\ell}\nu_{\ell}$ or $W^- \to
\ell\bar{\nu}_{\ell}$), ``leptonic decay''; or an up-type quark and a
down-type quark ($W^+ \to q \bar{q}'$ or $W^- \to \bar{q} q'$),
``hadronic decay''.  The result presented here uses the lepton+jets
decay channel (with $q\bar{q}'b\ell\bar{\nu_{\ell}}\bar{b}$ or
$\bar{\ell}\nu_{\ell}b\bar{q}q'\bar{b}$ in the final state), where one
of the two $W$~bosons decays leptonically into an electron or a muon, and
the other decays hadronically.  All the quarks in the final state evolve into
jets of hadrons.  Events with tau leptons are not selected directly,
but may contribute a few percent of the total sample via leptonic
cascade decays or fake jets.  The most recent $m_t$ measurements
obtained at the Tevatron using the lepton+jets topology are reported
in Ref.~\cite{TeVcomb}, while the results of an earlier version of the
present analysis using 955~pb$^{-1}$ of integrated luminosity are
reported in Ref.~\cite{meatprl}.  The distinctive feature of this
analysis is the use of matrix element calculations to describe the
dominant background contribution.  The result presented here uses a more
than three times larger data sample than the earlier version, and
employs a more detailed likelihood function.

The leptons and jets resulting from the top-antitop quark pair
($t\bar{t}$) decay are detected in the CDF~II general-purpose particle
detector that is described in detail elsewhere~\cite{CDFtdr}.
Azimuthally and forward-backward symmetric about the beam-line, the
detector contains a high precision particle tracking system immersed
in a 1.4 T magnetic field and surrounded by calorimetry, with muon
detectors on the outside. A right-handed spherical co-ordinate system
is employed, with the polar angle $\theta$ measured from the proton
beam direction, the azimuthal angle $\phi$ in the plane perpendicular
to the beam-line, and the distance $r$ from the center of the
detector. Transverse energy and momentum are defined as $\Et\equiv E
\sin\theta$ and $p_T\equiv p \sin\theta$, where $E$ and $p$ denote
energy and momentum. Pseudorapidity is defined as $\eta\equiv-\ln
\tan\left(\theta/2\right)$.

\begin{table}
\begin{center}
\begin{tabular}{llll}
\hline
Electron & $\Et$      &$> \rm 20~GeV$ & $|\eta|<1.1$ \\
or Muon & $p_T$       &$> \rm 20~GeV/{\it c}$ & $|\eta|<1.0$ \\
$\missEt$ & $\missEt$ &$> \rm 20~GeV$ & $|\eta|<3.6$ \\
Jets & $\Et$          &$> \rm 20~GeV$ & $|\eta|<2.0$ \\
\multicolumn{4}{l}{Four jets; at least one from a $b$~quark}\\ 
\hline
\end{tabular}
\end{center}
\caption{Event selection criteria.}
\label{tab:es}
\end{table}

This measurement makes use of CDF II data collected between February
2002 and August 2008, representing approximately \IntLum\ of integrated
luminosity. The event selection criteria (Table~\ref{tab:es}) are
tuned to select the lepton+jets final-state particles, requiring that
each event must have exactly one high-$\Et$ electron or high-$p_T$
muon, exactly four high-$\Et$ jets, and a significant amount of
missing $\Et$, $\missEt$~\cite{missEt}, characteristic of the
undetected neutrino. Jets are reconstructed using a cone
algorithm~\cite{coneAlgor}, with the cone radius $\Delta
R\equiv\sqrt{\left(\Delta\eta\right)^2+\left(\Delta\phi\right)^2}=0.4$.
At least one of the four jets must be identified as originating from a
$b$~quark via the {\sc SECVTX} algorithm~\cite{secvtx}, which detects
displaced secondary vertices characteristic of the decay of long-lived
$b$~hadrons.  A total of \NcResult\ events are selected, of which 76\%
are expected to be $t\bar t$ events (Table~\ref{tab:mc}). Of the 24\%
of events expected to be background, it is predicted that 69\% arise
from the production of a $W$~boson in conjunction with 4 jets
($W$+jets), 19\% come from multi-jet QCD production (non-$W$), while
the remaining 12\% are from sources such as diboson and
single-top-quark production. These fractions are estimated using
theoretical cross-sections, Monte Carlo (MC) simulated events, and
data.  The $t\bar t$ events are generated using the Lund Monte Carlo
program {\sc PYTHIA}~\cite{pythia}, with a top-quark mass of $\rm
175~GeV/{\it c}^2$ and a $t\bar t$ production cross section of
$6.7\pm0.8$~pb~\cite{crossSecNumber}.  The $W+$jets and $Z+$jets
events are generated using the {\sc ALPGEN} generator~\cite{alpgen}
while the single-top-quark events are generated using the {\sc
  MADEVENT} package~\cite{madEvent}, in both cases also using {\sc
  PYTHIA} to perform the parton showering and hadronization. Diboson
events are also generated using {\sc PYTHIA}. In addition, data are
used for non-$W$ events~\cite{evtSelect}.

\begin{table}
\begin{center}
\begin{tabular}{llcc}
sample& \# of events & \% of total & \% of bkg \\
\hline
\hline

$t\bar t$ signal & 425.0 $\pm$ 58.9 & 76.0\% & - \\
\hline
$W+$jets & 92.6 $\pm$ 15.9 & 16.6\% & 69.0\% \\
non-$W$ & 25.0 $\pm$ 12.5 & 4.5\% & 18.7\% \\
single top quark& 6.6 $\pm$ 0.4 & 1.2\% & 4.9\% \\
diboson  & 6.0 $\pm$ 0.6 & 1.1\% & 4.5\% \\
$Z+$jets & 3.9 $\pm$ 0.5 & 0.7\% & 2.9\% \\
\hline
total & 559.2 $\pm$ 67.0 & 100\% & - \\
Observed & \!\:\!\:\!\:\NcResult
\end{tabular}
\end{center}
\caption[Number of expected signal and background events.]{Number of
expected signal and background events, 
corresponding to the total integrated luminosity of \IntLum. 
The percentages are used when generating Monte Carlo 
simulated experiments. } 
\label{tab:mc}
\end{table}

This analysis employs an unbinned maximum likelihood method
\cite{meatprl,D0nature,WHflo}. The $m_t$-dependent probability density
function (p.d.f.)\ is calculated for each event in the data sample:

\begin{equation}
  \mathcal{P}(k) = \Cs P_{s}(k) + (1 - \Cs)
   P_{b}(k) \mathrm{,}
  \label{eqn:probability}
\end{equation}

\noindent where $k\equiv(E_i,\vec{p}_i)$ represents the measured
kinematic quantities of the event, $P_s$ and $P_b$ are respectively
the normalized p.d.f.s for signal and background events, and $\Cs$ is
the signal fraction parameter (constrained $0 \le \Cs \le 1$).  Signal
events are defined as events consistent with $q\bar q \rightarrow
t\bar t$ production and $t\bar t$ decay into the lepton+jets channel,
as described by the leading-order (LO) matrix element evaluated by
Mahlon and Parke~\cite{matrixElem}.  Background events are assumed to
be described by a matrix element for $W+$jets production, which is
calculated using a sum of 1286 $W+$4-partons amplitudes for 592
subprocesses encoded in the {\sc VECBOS} MC event
generator~\cite{vecbos}.  This approximation does mean that there are
some events that, in principle, are not described by either $P_s$ or
$P_b$, including non-$W$, single top, diboson, $Z+$jets, and
$W+bb+$2-partons events, as well as $W+$jets events from $W+$0-, 1-,
2- and 3-partons processes.  However, studies with MC simulated events
show that the ratio $P_b/P_s$ calculated for all of these event types
is similar to that for $W+$4-partons events, and that, in practice,
such events mostly contribute to the likelihood function via the $P_b$
term and do not add any more bias than the $W+$4-partons events or
than the poorly reconstructed $t\bar t$ events
themselves~\cite{JacobThesis}.  Any residual bias in the measured
top-quark mass is removed at the end, as described later in the paper.

The signal and background p.d.f.s, $P_s$ and $P_b$, are constructed in
analogous fashions, starting with the appropriately normalized
parton-level differential cross-section~\cite{pdg}, $d\hat{\sigma}_s$
or $d\hat{\sigma}_b$, which is then convolved with parton distribution
functions (PDFs) and a jet-to-parton transfer function
$W(k,\varkappa)$.  $P_s$ is thus given by

\begin{align}
 &P_{s}(k; m_t, \JES) = \frac{1}{n_{jp}}\sum_{\mathrm{jet\:perm.}}^{n_{jp}} \frac{1}{\hat{\sigma}_s(m_t)} \frac{1}{ A_s(m_t, \JES)} \times \label{eqn:prob4} \\ 
 &  \int d\hat{\sigma}_s(\varkappa; m_t) \;   dx_{\mathrm{Bj}}^1dx_{\mathrm{Bj}}^2
 \; W(k,\varkappa; \JES)  \;   f(x_{\mathrm{Bj}}^1) (x_{\mathrm{Bj}}^2) \nonumber  \mathrm{,}
 \end{align}

\noindent where $\varkappa\equiv(\varepsilon_i,\vec{\pi}_i)$
represents the actual event parton-level kinematic quantities
corresponding to the measured quantities $k$, and parameter $\JES$ is
defined in the next paragraph.  The PDFs $f(x_{\mathrm{Bj}})$ define
the probability density for a colliding parton to carry a longitudinal
momentum fraction $x_{\mathrm{Bj}}$ and are given by {\sc
  CTEQ5L}~\cite{cteq}.  $A_s$ is the mean acceptance function for
signal events, a normalization term that is the consequence of the
constriction of the phase-space of the integral by the event selection
cuts and by the detector acceptance.  The average over the jet
permutations, $n_{jp}$, is due to ambiguity in assigning final state
jets to partons.  The fact that the two light quarks in the final
state are indistinguishable allows the reduction from the original 24
permutations to 12 in the expression for $P_s$, and the $b$-tagging
information allows a further reduction to 6 assignments for events
with one identified $b$-jet and 2 for events with both $b$-jets
identified.  In the similar expression for $P_b$, all 24 permutations
are averaged.

The jet-to-parton transfer function $ W(k,\varkappa)$ is a
p.d.f.\ describing the probability density for an event with out-going
partons and charged lepton with $\varkappa$ to be measured as
reconstructed $k$. The charged lepton is assumed to be well-measured,
allowing the use of a Dirac $\delta$-function to represent the mapping
between its parton-level momentum, $\vec{\pi}_{\ell}$, and its
reconstructed momentum, $\vec{p}_{\ell}$.  For the four jets, the
function is obtained by parameterizing the jet-to-parton mapping
observed in fully simulated {\sc PYTHIA} $t\bar t$ events.  These
events contain all of the information about the original partons as
well as the measured jets.  The simulation includes physical effects,
such as radiation and hadronization, as well as the effects of
measurement resolution and of the jet reconstruction algorithm.  The
parameterization is made in two parts that are assumed to be
independent: the energy transfer function $\Wj$, describing the jet
energies $E$, and the angular transfer function $\Wa$, describing the
mapping for the jet angles.  The jet-to-parton transfer function is
thus given by

\begin{align}
  W(k,\varkappa; \JES)&=\delta^3(\vec{p}_{\ell}-\vec{\pi}_{\ell}) \: \Wa \label{eqn:TF}  \\
   &\times \prod_{i=1}^{4}{\left(  \frac{1}{E_i p_i}\Wj^i(E_i,\varepsilon_i; \JES) \right)}  \nonumber  \mathrm{.}
\end{align}

The reconstructed jet energies, $E_i$, used in the function $\Wj$ are
not just the raw calorimeter energy deposits, but are first calibrated
so that they represent the combined energies released in the
calorimeter by the many particles constituting each jet. This is
achieved using the CDF jet energy scale (JES)
calibration~\cite{JESnim}, which is subject to a significant
systematic uncertainty. The uncertainties of individual jet energy
measurements, $\sigma(E_i)$, are therefore correlated, and their
fractional JES uncertainty, $\sigma(E_i)/E_i$, is typically
$\sim3\%$. If this were included as a systematic uncertainty on the
measured $m_t$ it would reduce the measurement precision drastically;
in fact, each 1\% of fractional JES uncertainty would add about $\rm
1~GeV/{\it c}^2$ uncertainty to the measured $m_t$~\cite{JEScorrel}.
However, such a treatment overestimates the uncertainty because the
energies of the two daughter jets of the hadronically decaying
$W$~boson can be constrained based on the known $W$~boson mass.
Applying this constraint to all events in the data sample while
allowing the jet energies to be shifted results in the \textit{in
  situ} measurement of the JES correction, $\JES$, defined as the
number of $\sigma(E_i)$ values by which the energy of each jet is
shifted in the likelihood fit.  This effectively re-calibrates the
measured jet energies based on the known $W$~boson mass and replaces a
large component of the JES systematic uncertainty with a much smaller
statistical uncertainty on the $\JES$.  The $\JES$ dependence of the
jet energies is included in the parameterization of the function
$\Wj$.  This parameterization is made in eight bins in pseudorapidity
$|\eta|$, separately for light and $b$-jets, using a sum of two
Gaussians as a function of the difference between the parton energies
and the corrected jet energies as measured in a sample of {\sc PYTHIA} $t\bar t$ events that pass the same selection criteria as the data.

\begin{figure}
\begin{center}
\includegraphics[width=0.4\textwidth]{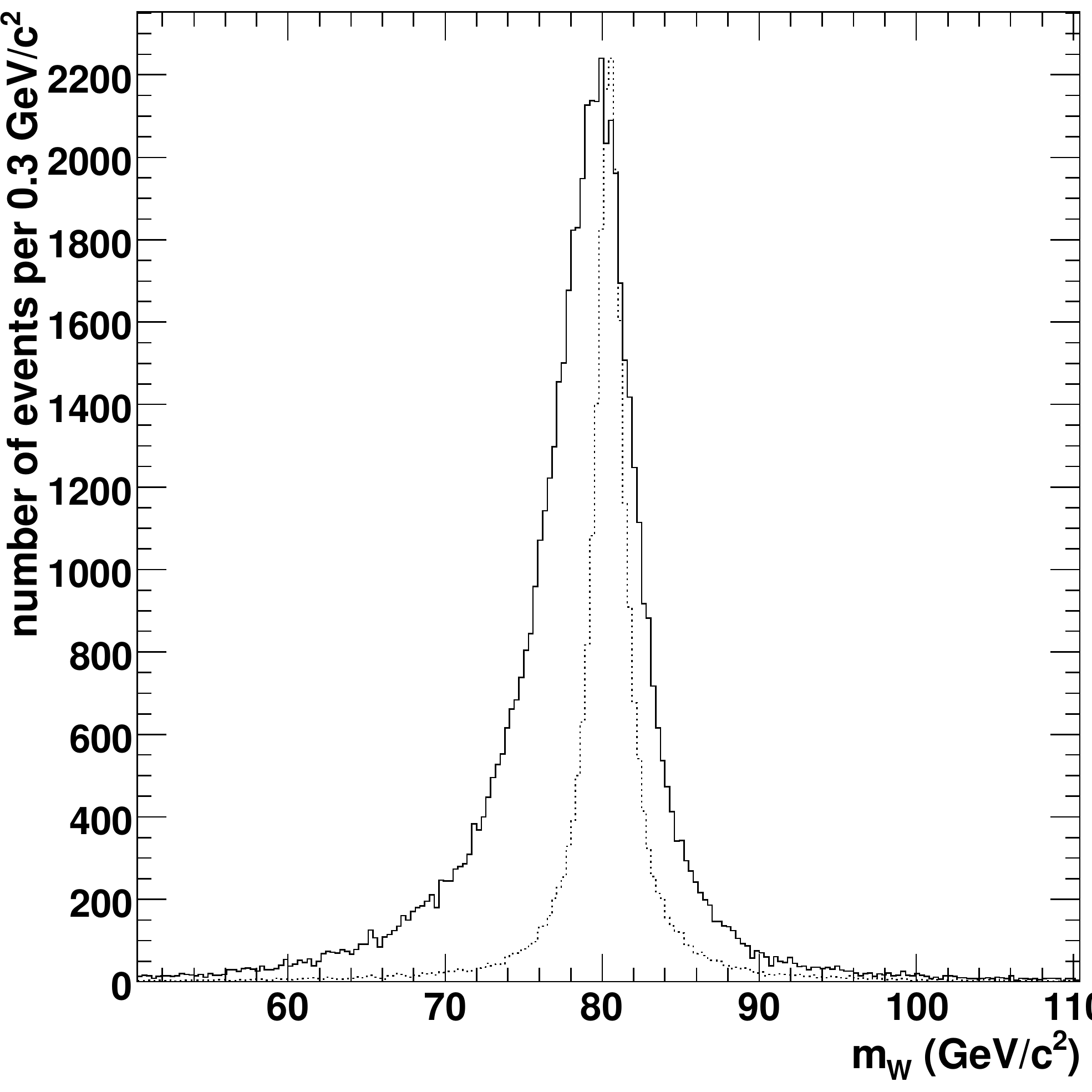}
\caption{ The reconstructed 2-jet invariant mass of the
hadronically decaying $W$ boson, $m_W$, 
for measured jet angles (solid line) and for parton-level angles (dotted line),
 obtained after assuming the primary parton energy as jet energy. 
For ease of comparison, 
the parton-level distribution is normalized 
so that the maxima of the two distributions are the same.
}\label{fig:breitw}
\end{center}
\end{figure}

In an earlier version of this analysis~\cite{meatprl}, the
jet-to-parton transfer functions for all jet angles were approximated
by Dirac $\delta$-functions.  The introduction of the function $\Wa$
was motivated by a discrepancy noticed in simulated $t \bar t$ events
in the 2-jet effective invariant mass of the hadronically decaying $W$
boson, $m_W$.  Even when the true simulated parton-level jet energies
are used, instead of the corresponding reconstructed detector-level
values, the use of the measured jet angles rather than their
parton-level values causes a significant shift of the reconstructed
$m_W$ from its nominal value, as illustrated in Fig.~\ref{fig:breitw}.

\begin{figure}
\begin{center}
\includegraphics[width=0.4\textwidth]{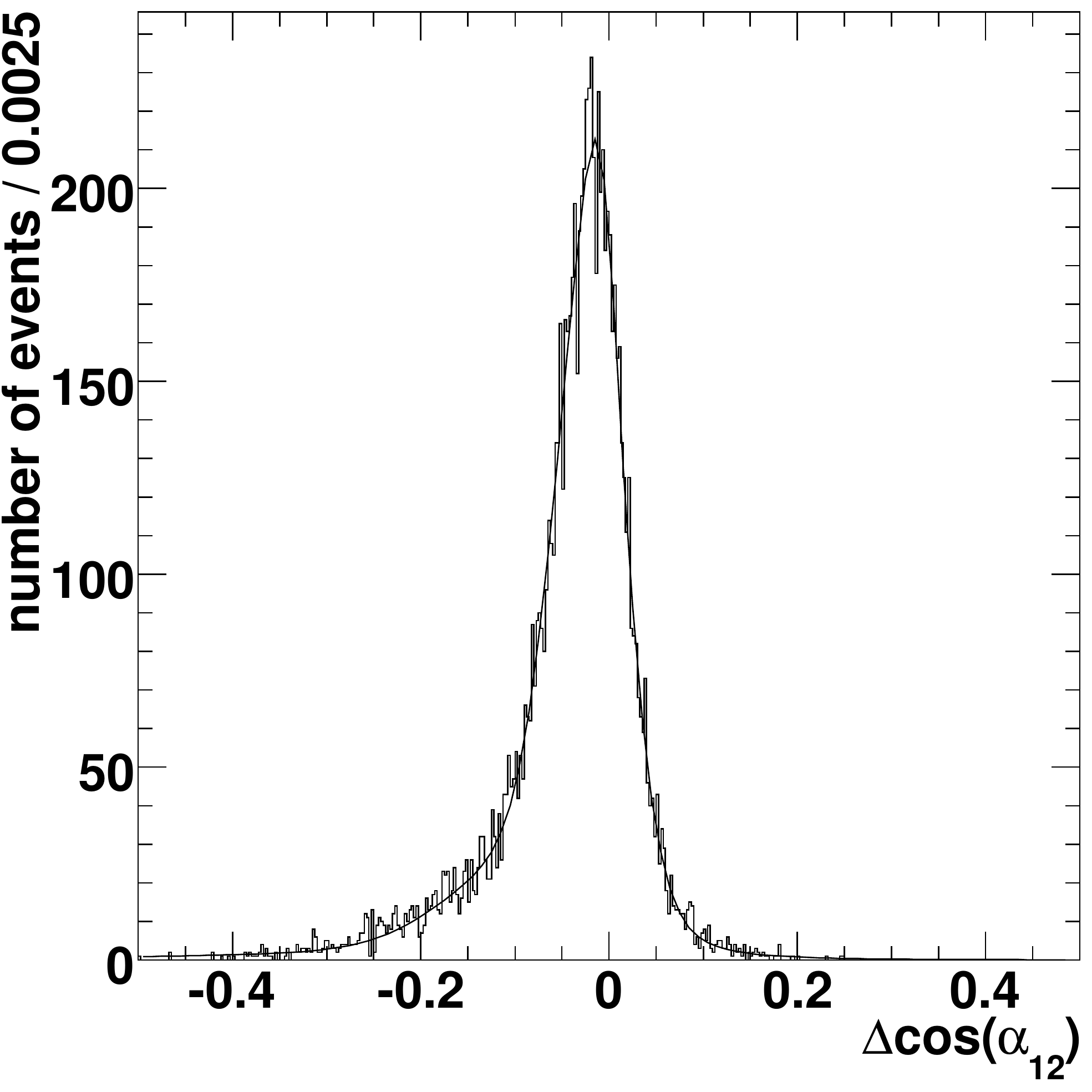}
\includegraphics[width=0.4\textwidth]{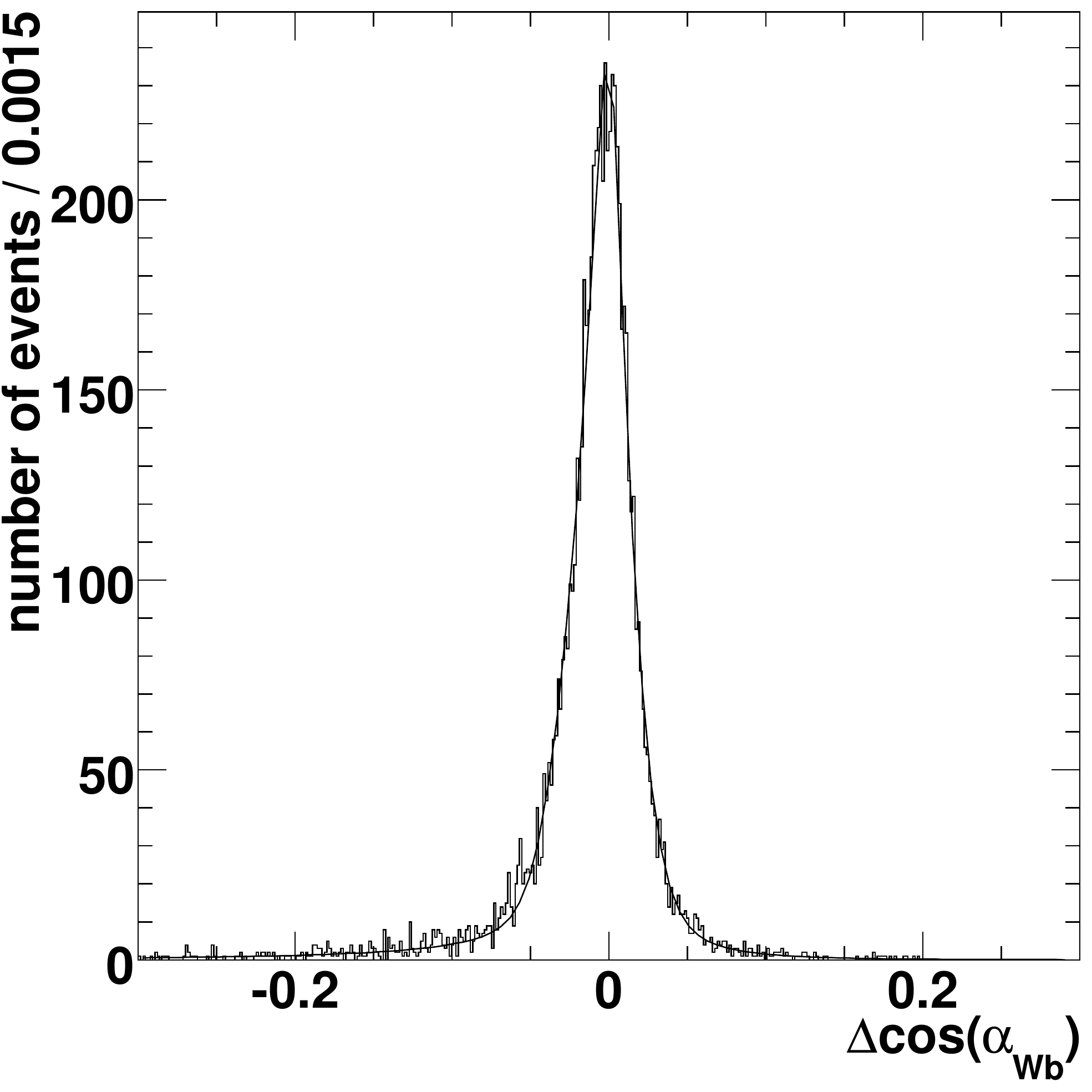}
\caption{Examples of parameterization of the functions $W^{12}_A$ 
and $W^{Wb}_A$ in  the bins where $0.2 < \cos{(\alpha_{12})} < 0.4$ 
and $0.2 < \cos{(\alpha_{Wb})} < 0.4$.
The histograms show MC simulation events and the curves represent the parameterization.
}\label{fig:Wangular}
\end{center}
\end{figure}

There is also a negative skewness in the distribution for measured
angles, and since parton-level jet energies are used, the observed
effects are due to the differences between the measured angles and the
parton-level angles alone.  The peak of the $m_W$ distribution, when
fit by a Breit-Wigner distribution, corresponds to a $W$~boson pole
mass of $\rm 79.5~GeV/{\it c}^2$, a $\rm -0.9~GeV/{\it c}^2$ shift
from its parton-level value of $\rm 80.4~GeV/{\it c}^2$.  This is
found to be a result of a correlation between the measured jet
directions: the measured angle, $\alpha_{12}$, between the two jets
is, on average, reduced so that the two jets appear closer together
than their parent partons, which can be seen in Fig.~\ref{fig:Wangular}.  Since
the apparent $W$~boson mass is utilized to measure $\JES$ and thus
calibrate the measured jet energies, a jet-to-parton transfer function
describing the change in the angle $\alpha_{12}$ is important in
making an accurate measurement of $\JES$ and thus the top-quark mass.
The function $\Wa$ also describes a much smaller correlation effect
seen in the angle $\alpha_{Wb}$ between the hadronic-side $b$-jet and
the hadronically decaying $W$~boson.  The function $\Wa$ is thus
parameterized using two different functions, $W^{12}_A$ and
$W^{Wb}_A$, describing the mappings for the angles $\alpha_{12}$ and
$\alpha_{Wb}$.  The remaining angles describe resolution effects
rather than the correlations and, due to computational constraints,
are assumed to be well measured with their contributions to $W_A$
approximated by Dirac delta functions.

The functions $W^{12}_A$ and $W^{Wb}_A$ are both fit using a sum of a
skew-Cauchy distribution and two Gaussians, describing the change in
the cosine of the relevant angle, $\Delta \cos{ (\alpha_{12})}$ and
$\Delta \cos{ (\alpha_{Wb})}$, from partons to measured jets.  Since
the correlation effects are stronger in jets that are closer together,
the functions are parameterized in bins of $\cos(\alpha_{12})$ and
$\cos(\alpha_{Wb})$, respectively; one example for each function is
shown in Fig.~\ref{fig:Wangular}.

\begin{figure}
\begin{center}
\includegraphics[width=0.4\textwidth]{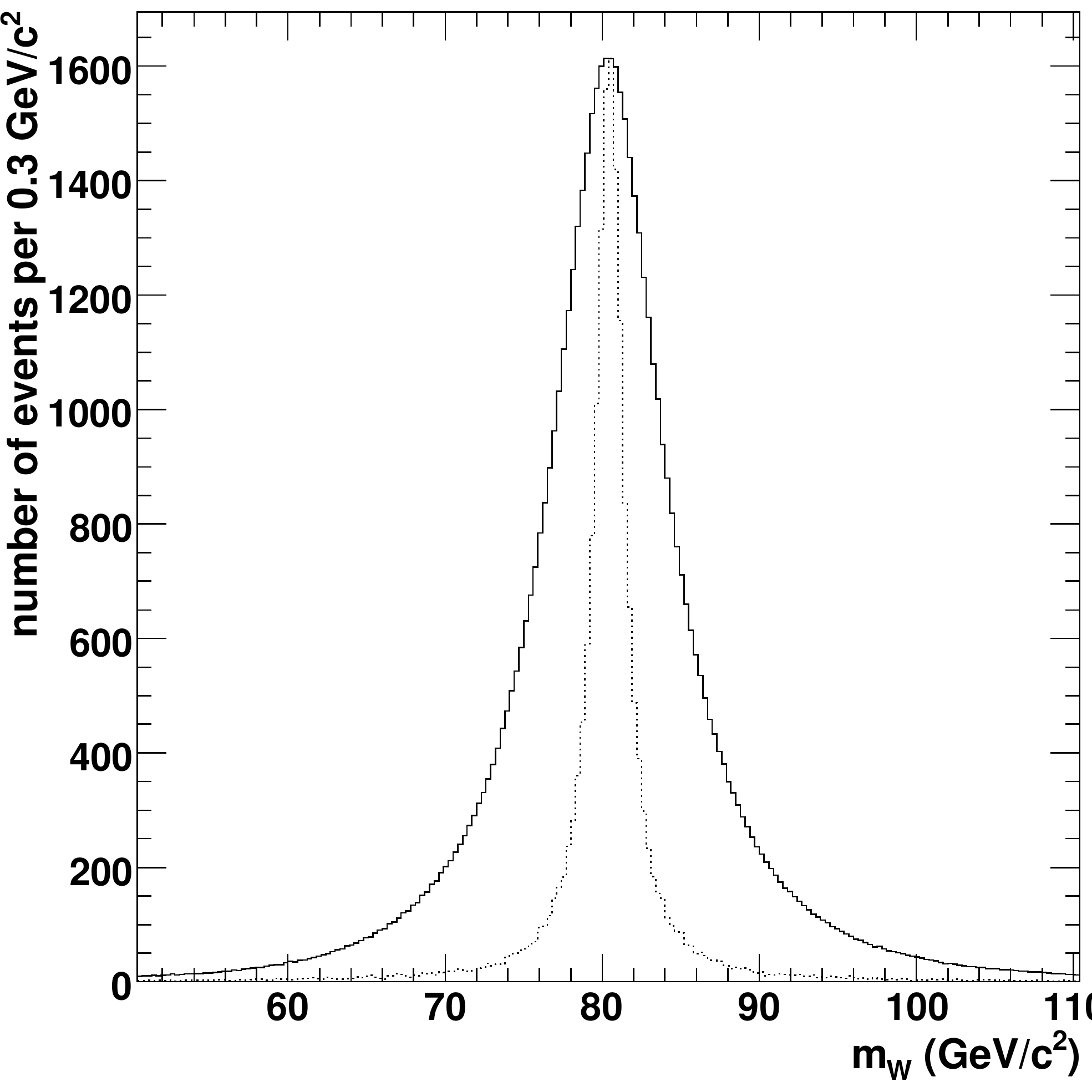}
\caption{ The $m_W$ distribution for measured angles from Fig.~\ref{fig:breitw} 
is plotted (solid line) after 
convolution with the function $\Wa$.
For ease of comparison, 
the parton-level distribution (dotted line) is normalized
so that the maxima of the two distributions are the same.
}\label{fig:breitw_conv}
\end{center}
\end{figure}

The $m_W$ distribution after convolution with the function $\Wa$ is
shown in Fig.~\ref{fig:breitw_conv}.  The skewness is removed and the
mean value agrees well with the parton-level distribution.

The 20 integration variables (3 for each final-state particle and the
$x_{\mathrm{Bj}}$ for each initial state parton, assuming zero
transverse momentum for the $t\bar t$ pair) in the expression for the
signal and background p.d.f.s (Eq.~\ref{eqn:prob4}) are reduced to 16
by integrating over the 4-momentum conservation Dirac
$\delta$-function inherent in the expression for
$d\hat{\sigma}_s$. The charged lepton 3-momentum integration and all
but two of the jet angular integrations are made trivial by the Dirac
$\delta$-functions in the function $W(k,\varkappa)$, leaving 7
integration variables.  In $P_s$, this is further reduced to 5
variables via a change of variables to the squared masses of the top
quarks and by using the narrow-width approximation for the
Breit-Wigner distributions of both top-quark decays in the $t\bar t$
matrix element.  The integral is then evaluated using the {\sc
  VEGAS}~\cite{vegas} adaptive Monte Carlo integration
algorithm~\cite{meatprl}, which uses importance sampling, which means
that the sample points are concentrated in the regions that make the
largest contribution to the integral.

The treatment of $P_b$ is unchanged since the previous version of this
analysis~\cite{meatprl}, except for the updated energy transfer
function $\Wj$.  The integrand in the expression for $P_b$ is much
more computationally intensive than for $P_s$ and a simplified Monte
Carlo method of integration is employed, giving reasonable convergence
with an execution time comparable to that of $P_s$.  The
simplifications used in this computation of $P_b$ include setting the
function $\Wa$ to a Dirac $\delta$-function for all angles, using a
narrow width approximation for the $W$~boson decay, and neglecting the
$\JES$ dependence of the function $\Wj$.  Therefore, the value of
$P_b$ for each event does not depend on the likelihood parameters
$m_t$ and $\JES$, while $P_s$ is a two-dimensional function of those
parameters~\cite{meatprl}.  In this approximation, the product of the
background p.d.f.  normalization terms (corresponding to the variables
$\hat{\sigma}_b \cdot A_b$ in Eq.~\ref{eqn:prob4}) is set to a
constant, whose value is chosen to optimize the statistical
sensitivity of the method, effectively providing an appropriate
relative normalization with respect to $P_s$.

The log-likelihood function is given 
as a sum over the 578 events in the sample:

\begin{align}
  \ln \, & \mathcal{L}(k;m_t,\JES,\Cs) =  \label{eqn:loglike} \\   
    & \sum^{578}_{i=1} \ln \left[ \Cs {P_{s}(k_i;m_t,\JES)} + (1-\Cs) P_{b}(k_i) \right] \nonumber
\mathrm{.}  
\end{align}

\begin{figure}
\begin{center}
\includegraphics[width=0.45\textwidth]{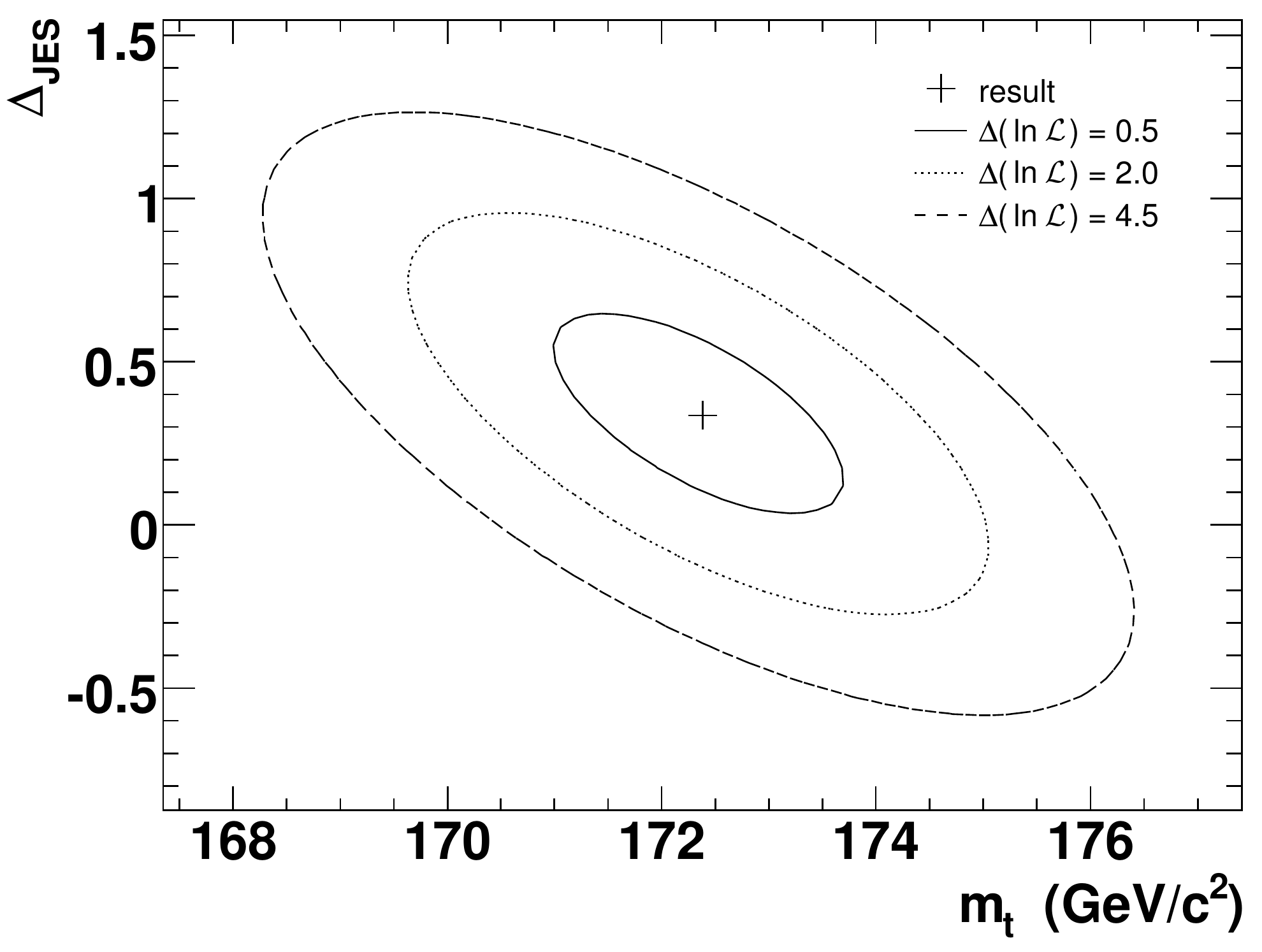}
\caption[The measurement result.]{The measurement result and the
contour ellipses of the parabolic fit corresponding to the one-, two- and 
three-$\sigma$ confidence intervals for the statistical uncertainty on $m_t$ and $\JES$.}\label{f:Result}
\end{center}
\end{figure}

\noindent It is calculated on a two-dimensional $31\times17$ grid in
$m_t$ and $\JES$, spanning $145 \leq m_t \leq \rm 205~GeV/{\it c}^2$
and $-4.8 \leq \JES \leq 4.8$, with a spacing between grid points of
$\rm 2~GeV/{\it c}^2$ in $m_t$ and 0.6 in $\JES$.  To optimize
computational time, the bin size is chosen to be as large as possible
without appreciably affecting the fit result.  The third likelihood
parameter, the signal fraction parameter $\Cs$, is allowed to vary
continuously (within the constraint $0 \le \Cs \le 1$), and the
likelihood function is maximized with respect to $\Cs$ at each point
on the grid using the {\sc MINUIT} program~\cite{minuit}. The
resulting surface described on the grid is the profile
log-likelihood, maximized for $\Cs$.  The top-quark mass, $m_t$, and the jet
energy scale correction, $\JES$, are measured by making a
two-dimensional parabolic fit to the surface, consistent with the
expectation for the likelihood function to be Gaussian near its
maximum. The maximum of the parabola gives the measured $m_t$ and
$\JES$, while the measured $\Cs$ is taken from its value at the grid
point of maximum likelihood. The estimated one-$\sigma$ statistical
uncertainty of the measurement is represented by the ellipse
corresponding to a change in log-likelihood $\Delta
\ln\mathcal{L}=0.5$ from the maximum of the fitted parabola. The
values of $m_t$ and $\JES$ are anti-correlated
(Fig.~\ref{f:Result}). No correlation is observed between $\Cs$ and
$m_t$ or $\JES$.

%
%
The accuracy of the measured $m_t$ and $\JES$, and their
uncertainties, are checked using ensembles of MC simulated
experiments, using the MC samples previously mentioned with the
addition of 22 $t\bar t$ samples generated with values of $m_t$
between $161$ and $\rm 185~GeV/{\it c}^2$. The numbers of $t\bar t$
events and those of the various backgrounds are Poisson fluctuated
around the values shown in Table~\ref{tab:mc}.  Studies of the
relationships between the known input simulation parameters and their
corresponding measurements show no evidence of bias when a clean 
sample of MC simulated $t \bar t$ events is used, containing only lepton+jets events with
correct jet-parton matching.  However, the presence of signal
events with jets which are poorly or incorrectly matched to partons
and events which do not match the decay hypothesis biases the
likelihood fit result and increases the pull width.  The presence of
background events also biases the fit, due to the backgrounds that are
not well described by $P_b$ and the approximations in $P_b$.  The bias
is removed using a set of functions obtained from a fit to the MC
simulation and parameterized in terms of the measured $\JES$
and $\Cs$~\cite{JacobThesis}. This amounts to adding $\rm 1.1~GeV/{\it
  c}^2$ to the $m_t$ value produced by the likelihood fit and
multiplying the uncertainty by 1.26 so that the pull width is
consistent with unity.  The systematic uncertainty due to this
 measurement calibration is small, as shown in Table~\ref{tab:syst}.
%
%
%

Despite the reduction from the \textit{in situ} $\Delta_{\rm JES}$
calibration, the remaining uncertainty from JES obtained by varying the parameters in JES~\cite{JESnim} is among the 
largest systematic uncertainties of the measurement (Table~\ref{tab:syst}). Other significant systematic
uncertainties are mainly a result of assumptions made in the
simulation of the events that are used in the tuning and calibration
of the measurement method.  In most cases, they are evaluated by
varying different aspects of the MC simulation, such as signal MC
generator ({\sc PYTHIA} versus {\sc HERWIG}~\cite{herwig}), color
reconnection model tune (Apro versus
ACRpro~\cite{CRpaper,CRtalk,CRtunes}), and parameters of initial and final
state radiation (ISR and FSR).  A detailed description of the systematic effects has been published
elsewhere~\cite{mtmPRD}.  The systematic uncertainties for each effect
are added in quadrature, resulting in a total estimated systematic
uncertainty of $\rm 1.3~GeV/{\it c}^2$ (Table~\ref{tab:syst}).

\begin{table}
\begin{center}
\begin{tabular}{cc}
Systematic     & (GeV/{\it c}$^2$) \\ \hline
\hline
MC Generator  & 0.70 \\
Residual JES  & 0.65 \\
Color Reconnection  & 0.56 \\
$b$-jet energy         & 0.39 \\
Background         & 0.45 \\
ISR and FSR         & 0.23 \\
Multiple Hadron Interactions & 0.22 \\
PDFs           & 0.13 \\
Lepton Energy & 0.12 \\
Measurement Calibration & 0.12 \\
\hline
Total         & 1.31 \\
\end{tabular}
\end{center}
\caption{\label{tab:syst}Contributions to the total expected systematic uncertainty.}
\vspace{1cm}
\end{table}

The measurement is made using the data sample of \NcResult\ events, yielding
\begin{align}
	m_t &= 172.4 \rm \pm 1.4\:(stat\mathrm{+}\JES) \pm
	1.3\:(sys)~GeV/{\it c}^2 \nonumber \\
	m_t &= 172.4 \rm \pm 1.9\:(total)~GeV/{\it c}^2 \label{e:resulttotal} 
	\mathrm{,}
\end{align}

\noindent with $\rm \JES=0.3 \pm 0.3\:(stat)$.  The central value and
the contour ellipses corresponding to the one-, two- and
three-$\sigma$ statistical confidence intervals 
of the measurement are illustrated in
Fig.~\ref{f:Result}.  The overall statistical uncertainty on the
measured top-quark mass is labeled ``stat+$\JES$'' because it includes
the uncertainty on $m_t$ due to the statistical uncertainty on the
measured $\JES$, i.e.\ the uncertainty is given by half of the full
width of the one-$\sigma$ contour of Fig.~\ref{f:Result}.

In conclusion, a precise measurement of the top-quark mass has been
presented using CDF lepton+jets candidate events corresponding to an
integrated luminosity of \IntLum.  Using an improved matrix element
method with an \textit{in situ} jet energy calibration, the top quark mass is measured to be $m_t
= \rm 172.4 \pm 1.9~GeV/{\it c}^2$.

\begin{acknowledgments}
We thank the Fermilab staff and the technical staffs of the
participating institutions for their vital contributions. This work
was supported by the U.S. Department of Energy and National Science
Foundation; the Italian Istituto Nazionale di Fisica Nucleare; the
Ministry of Education, Culture, Sports, Science and Technology of
Japan; the Natural Sciences and Engineering Research Council of
Canada; the National Science Council of the Republic of China; the
Swiss National Science Foundation; the A.P. Sloan Foundation; the
Bundesministerium f\"ur Bildung und Forschung, Germany; the Korean
World Class University Program, the National Research Foundation of
Korea; the Science and Technology Facilities Council and the Royal
Society, UK; the Institut National de Physique Nucleaire et Physique
des Particules/CNRS; the Russian Foundation for Basic Research; the
Ministerio de Ciencia e Innovaci\'{o}n, and Programa
Consolider-Ingenio 2010, Spain; the Slovak R\&D Agency; the Academy of
Finland; and the Australian Research Council (ARC).
\end{acknowledgments}

\bibliography{meatv3_prd}

\end{document}